%% file: main.tex
\documentclass[11pt,a4paper]{llncs}
\usepackage{amsmath, graphicx}
\usepackage[ruled,linesnumbered]{algorithm2e}
\usepackage{algorithmic}
\usepackage{paralist} % also compact lists
\usepackage{color}
\usepackage{multirow}
\usepackage{rotating}
\usepackage{graphicx,wrapfig,lipsum}
\usepackage{hyperref}
\usepackage[mathscr]{eucal} %% For \mathscr
\usepackage{amsbsy} %% For \boldsymbol
\usepackage{subfigure}
\usepackage{booktabs}

\usepackage{xcolor}
\usepackage{tikz}
\usetikzlibrary{snakes}
\usepackage[outercaption]{sidecap}  
\usepackage{multirow}
\usepackage{color}
\usepackage{epstopdf}
\usepackage{balance}
\usepackage{tablefootnote}
\usepackage{empheq} % added by christos - makes nice box for equations
\usepackage{moresize}
\usepackage{graphicx}

\usepackage{enumitem}
\setlist{nolistsep}
\usepackage{mdframed}
\usepackage{stackengine} 
\usepackage{sidecap}

\newcommand{\mourmeth}{\text{CommLLM}}
\newcommand{\newmethod}{$\mourmeth$\xspace}

\input{dfn}

\usepackage{marvosym}
\usepackage{fancyhdr}

\usepackage{geometry}
\newcommand{\keywords}[1]{\par\addvspace\baselineskip
\noindent\keywordname\enspace\ignorespaces#1}

\begin{document}
 
\title{LLMs Between the Nodes: Community Discovery Beyond Vectors}

\author{\large{Ekta Gujral  \and Apurva Sinha}}
\institute{egujr001@ucr.edu\\apurvasinha2003@gmail.com }
\authorrunning{\LARGE{LLMs Between the Nodes}}

\maketitle
\input{000abstract} 

\input{010introduction}
\input{020relatedwork}
\input{030method}

\input{040experiments}
\input{050conclusions}

\clearpage
\bibliographystyle{plain}
\bibliography{BIB/refs}

\vspace{2cm}

\section*{Authors}
\noindent {\bf Ekta Gujral} is staff data scientist in Walmart. She received Ph.D. in Computer Science from the University of California Riverside. Her research work focused on modeling and mining multi-aspect graphs with scalable streaming tensor decomposition for real-world applications. During her Ph.D. journey, She interned with Adobe, Snapchat, and LLNL Research teams. Her work spans a wide range of areas including tensor analysis, anomaly detection, explainable AI, and graph mining. She is motivated to build useful models that would deliver a great impact on society and business.\\

\noindent {\bf Apurva Sinha} is staff data scientist in Walmart. She received Master of Sciecne from University of Texas - Dallas. She did Bachelor of Science in Computer Science from Kalinga Institute of Industrial Technology, India. She has diverse expertise in advanced data analytics, data science, software development, Agile project, program management. Her research area includes machine learning, data analysis, NLP, and adaptive signal processing, with a vision to advance efficient and trustworthy artificial intelligence.\\
\end{document}

%% file: dfn.tex
\newcommand{\hide}[1]{}

\newcommand{\bit}{\begin{compactitem}}
\newcommand{\eit}{\end{compactitem}}
\newcommand{\ben}{\begin{compactenum}}
\newcommand{\een}{\end{compactenum}}

%% file: 000abstract.tex
\begin{abstract}
Community detection in social network graphs plays a vital role in uncovering group dynamics, influence pathways, and the spread of information. Traditional methods focus primarily on graph structural properties, but recent advancements in Large Language Models (LLMs) open up new avenues for integrating semantic and contextual information into this task. In this paper, we present a detailed investigation into how various LLM-based approaches perform in identifying communities within social graphs. We introduce a two-step framework called \newmethod, which leverages the GPT-4o model along with prompt-based reasoning to fuse language model outputs with graph structure. Evaluations are conducted on six real-world social network datasets, measuring performance using key metrics such as Normalized Mutual Information (NMI), Adjusted Rand Index (ARI), Variation of Information (VOI), and cluster purity. Our findings reveal that LLMs, particularly when guided by graph-aware strategies, can be successfully applied to community detection tasks in small to medium-sized graphs. We observe that the integration of instruction-tuned models and carefully engineered prompts significantly improves the accuracy and coherence of detected communities. These insights not only highlight the potential of LLMs in graph-based research but also underscore the importance of tailoring model interactions to the specific structure of graph data.
\end{abstract}

%% file: 010introduction.tex
\section{Introduction}
\label{sec:intro}
Community detection in social networks is crucial for understanding the underlying structure \cite{gujral2020beyond} and dynamics of complex systems. It helps uncover groups of nodes with dense internal connections, revealing patterns such as shared interests, behaviors, or functions. This technique is widely applied across domains like marketing (to target consumer groups), cyber security (to detect malicious clusters), biology (to identify functional modules in protein networks), and politics (to map ideological communities). As networks \cite{gujral2019hacd,al2018t} grow more intricate, traditional methods struggle with scalability and adaptability. Recently, large language models (LLMs) \cite{gu2024survey,dam2024complete,wei2025plangenllms} have emerged as powerful tools in this field. Their ability to reason over text and structure enables them to detect nuanced relationships and community patterns in both textual and graph data. LLMs are increasingly being integrated into research workflows, outperforming many classical algorithms in terms of flexibility, interpretability, and cross-domain applicability.

Despite their impressive capabilities across a wide range of natural language tasks, existing large language models (LLMs) encounter significant challenges when applied to community detection within social network graphs. One of the primary limitations is their lack of innate access to graph-structured data; LLMs are fundamentally designed to process sequential text rather than inherently parallel or interconnected structures such as graphs \cite{bommasani2021opportunities}. This sequential bias limits their ability to model complex topological features, such as node centrality, clustering coefficients, and higher-order neighborhood interactions, all of which are critical for accurately identifying communities. Moreover, LLMs do not naturally encode graph-specific inductive biases—such as permutation invariance or localized neighborhood aggregation—which are essential for tasks like node clustering and structural role identification. Their inability to capture hierarchical or multi-scale structures within graphs often results in shallow representations that fail to generalize across diverse network topologies. While techniques such as graph embeddings or hybrid models attempt to bridge this gap, these approaches typically decouple the graph topology from the language modeling process, which can lead to a loss of structural fidelity and semantic nuance.

Scalability is another core issue. LLMs, especially those deployed on large-scale or dynamic social networks, suffer from computational inefficiencies and memory bottlenecks. Representing large graphs in a form consumable by LLMs (e.g., through serialization or textual descriptions) becomes increasingly infeasible as network size grows. Additionally, LLMs lack temporal awareness and dynamic reasoning capabilities required to model evolving communities or time-sensitive interactions in dynamic social graphs. Furthermore, most LLMs are trained on general-purpose corpora and lack task-specific fine-tuning for graph mining or community detection tasks \cite{wang2023llm4graph}. As a result, they often struggle with domain adaptation, interpretability, and robustness when applied to network analysis. They may also be susceptible to spurious correlations or biases embedded in the training data, which can skew the understanding of social graph dynamics and community boundaries. Lastly, LLMs typically do not incorporate explicit constraints or priors related to graph theory, such as modularity maximization or spectral properties, which are central to many traditional community detection algorithms.
\begin{figure*}
	\begin{center}
	    \includegraphics[clip,trim=1cm 3cm 0cm 3cm,width=0.95\textwidth]{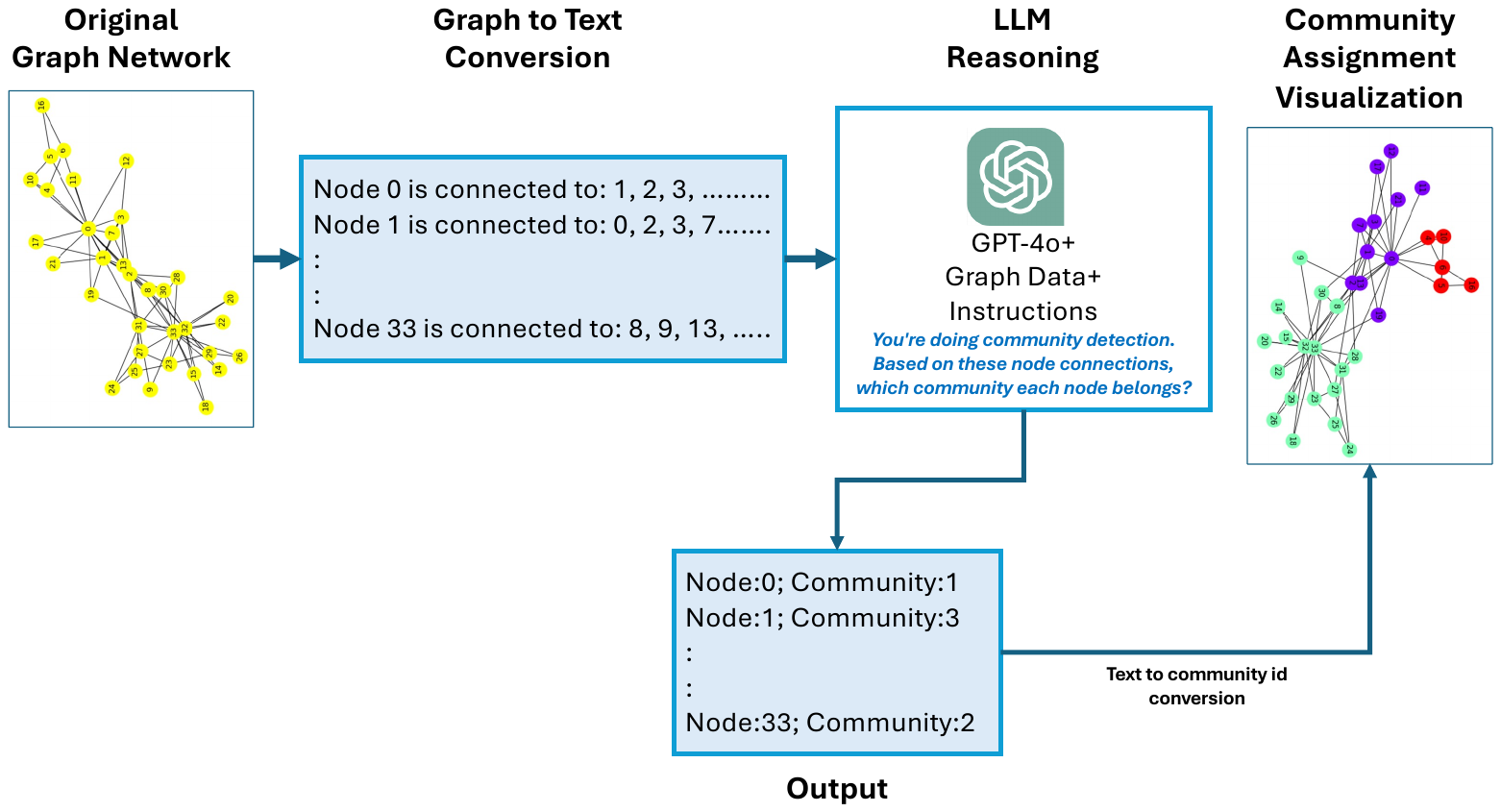}
		\caption{Illustration of \newmethod that includes two main steps: graph-to-text conversion and then LLM Reasoning to detect the communities in the graph} 
		\label{fig:mainimage}
	\end{center}
 \vspace{-0.2in}
\end{figure*}

In contrast, our work introduces a novel approach \newmethod (as shown in Figure \ref{fig:mainimage}) powered by GPT-4o \cite{openai2024gpt4o} that empowers large language models (LLMs) to detect communities more effectively by integrating graph-aware text representations with language-based reasoning. By embedding structural information directly into the input, our method preserves the topology of the network while leveraging the LLM’s capacity for pattern recognition and inference. This hybrid strategy allows the model to understand both the semantic and relational aspects of nodes. It enables nuanced community detection, even in complex or sparsely connected graphs. Additionally, our approach improves interpretability and generalization across different types of networks. Our contributions are summarized as follows:
\begin{itemize}
    \item We propose method \newmethod, which uses GPT-4o to detect communities in social network graphs. It includes two main steps: graph-to-text conversion and then prompt engineering to detect the communities in the graph. (See Figure \ref{fig:mainimage})
 \item The results on real datasets indicate that \newmethod outperforms the baseline LLM methods in context to community detection. Furthermore, prompt comparison experiments reveal that the designed prompt effectively aids LLMs in understanding the graph communities.
\item The dataset used and graph-to-text conversion script in this study is publicly available at link\footnote{\url{https://drive.google.com/drive/folders/1IKV-Qi_oz0hjAJHqCszLIbwlACmgNWXM}} for better reproducibility of the research.
\end{itemize}

%% file: 020relatedwork.tex
\section{Literature Review}
\label{sec:intro}
\subsection{Traditional Community detection}
Community detection in complex networks has garnered significant attention, leading to the development of a wide range of methodologies. Modularity-based approaches, such as the Louvain algorithm \cite{blondel2008fast}, aim to maximize a modularity score to identify groups of nodes with dense intra-community connections relative to a null model. Spectral clustering techniques utilize the eigenvectors of graph Laplacians to reveal partitions in the network structure \cite{newman2006modularity}. Label propagation algorithms \cite{raghavan2007near} iteratively update node labels based on neighbor consensus, offering scalability and simplicity. Statistical inference methods, including the Stochastic Block Model \cite{holland1983stochastic}, model networks probabilistically, often capturing assortative and disassortative structures. Hierarchical clustering builds dendrograms of nodes through agglomerative or divisive strategies, while edge betweenness-based methods \cite{girvan2002community} sequentially remove edges with high centrality to expose community boundaries. Random walk-based techniques, such as Walktrap, cluster nodes based on transition probabilities, whereas clique percolation methods identify overlapping communities through adjacent k-cliques. Information-theoretic frameworks like Infomap \cite{rosvall2008maps} aim to compress the description of network flows. More recent approaches include graph embedding techniques (e.g., DeepWalk \cite{perozzi2014deepwalk,al2020t}), which learn low-dimensional node representations for clustering, and deep learning-based models like graph neural networks (GNNs) that jointly learn features and community assignments \cite{kipf2016variational}. Additionally, matrix factorization methods (e.g., NMF) decompose adjacency matrices to detect latent community structure. Bayesian methods incorporate prior distributions into community inference \cite{peixoto2014hierarchical}, while evolutionary algorithms apply genetic operations for optimization. Tensor decomposition-based community detection \cite{gujral2018smacd,fang2024mntd} leverages multi-dimensional data representations to uncover complex community structures in networks. By decomposing high-order tensors (e.g., user-item-time) into latent factors, it captures hidden relationships and overlapping communities \cite{sheikholeslami2018identification,gorovits2018larc} that traditional methods might miss. Multi-resolution methods \cite{gujral2019hacd} further enable detection of communities at varying scales, addressing limitations related to resolution limits in modularity-based techniques.
\subsection{LLMs in graph reasoning}
In recent years, researchers have increasingly investigated the use of large language models (LLMs) for graph reasoning tasks \cite{wang2024graphtool}. The paper \cite{wang2023can} highlighted LLMs’ capabilities in graph reasoning and introduced the NLGraph dataset to assess these skills. The paper \cite{guo2023gpt4graph} further underscored the importance of graph encoding strategies in shaping how LLMs interpret and generate outputs. To facilitate LLM-based graph understanding, encoding methods such as Adjacency and Incident \cite{fatemi2023talk} have been developed, using integer-based node encoding with differing edge encoding schemes. While these methods mainly encode graph topology, they often fall short in addressing task-specific requirements like community detection. To bridge this gap, we propose a novel encoding method inspired by Incident, specifically tailored for the community detection domain by integrating community-aware structural cues. Additionally, although task-oriented prompts—like the Chain-of-Thought \cite{wei2022chain} and Build-a-Graph \cite{wang2023can} have been proposed to enhance LLM performance in graph reasoning, they lack the domain-specific context needed for community detection. Recently, the paper \cite{ni2024comgpt} introduced detecting community structure with large language models. Our work provide how different LLM models perform on the same task and it bridges the current gap in LLMs applicability to the community detection for social network graphs.

%% file: 030method.tex
\section{Approach}
\label{sec:method}
Traditional community detection algorithms, such as modularity maximization, spectral clustering, or label propagation, often rely on graph-theoretic computations. In contrast, we explore a novel approach \newmethod (Figure \ref{fig:mainimage}) by leveraging the reasoning capabilities of a large language model (LLM), specifically a GPT-4o, to infer community assignment directly from graph connectivity data.

Let $G=(V,E)$ be an undirected graph, where  $V=\{0,1,2,\dots,n-1\}$ is the set of nodes and $E \subseteq  V \times V$ is the set of edges representing connections between nodes. Each node is associated with a set of neighboring nodes $N(i)$, defined as:
$$N(i) = \{ j \in V \mid (i, j) \in E \}$$

Our goal is to partition the graph into a set of communities $C=\{C_1 ,C_2, C_3 \dots C_N\}$, such that nodes within the same community are more densely connected to each other than to nodes in other communities. Mathematically as follow:
$$\bigcup_{\ell=1}^k C_\ell = V, \quad C_i \cap C_j = \emptyset \quad \text{for } i \ne j$$
\subsection{Graph-to-Text Conversion}
\label{sec:method_GT}
Graph-to-text conversion is a fundamental task in natural language generation that entails the translation of structured graph data into coherent, informative, and contextually rich natural language descriptions. This process is essential for interpreting and communicating complex relational information embedded within graphs, facilitating a deeper understanding of the network’s structure and the interactions between its components. The goal is to generate textual summaries that not only describe the raw data but also convey the underlying patterns, behaviors, and trends represented by the graph.

To achieve this, we first extract salient features from the graph, such as node roles, edge weights, community affiliations, and subgraph structures. These elements serve as the foundational data for our approach, \newmethod, and are carefully analyzed to capture the most relevant and informative aspects of the graph’s topology. By encoding these features in a way that maintains the integrity of the graph’s structure, we can produce textual summaries that highlight key relationships and important dynamics within the network.

\begin{figure}
	\begin{center}
	    \includegraphics[clip,trim=2cm 3cm 2cm 3cm,width=0.85\textwidth]{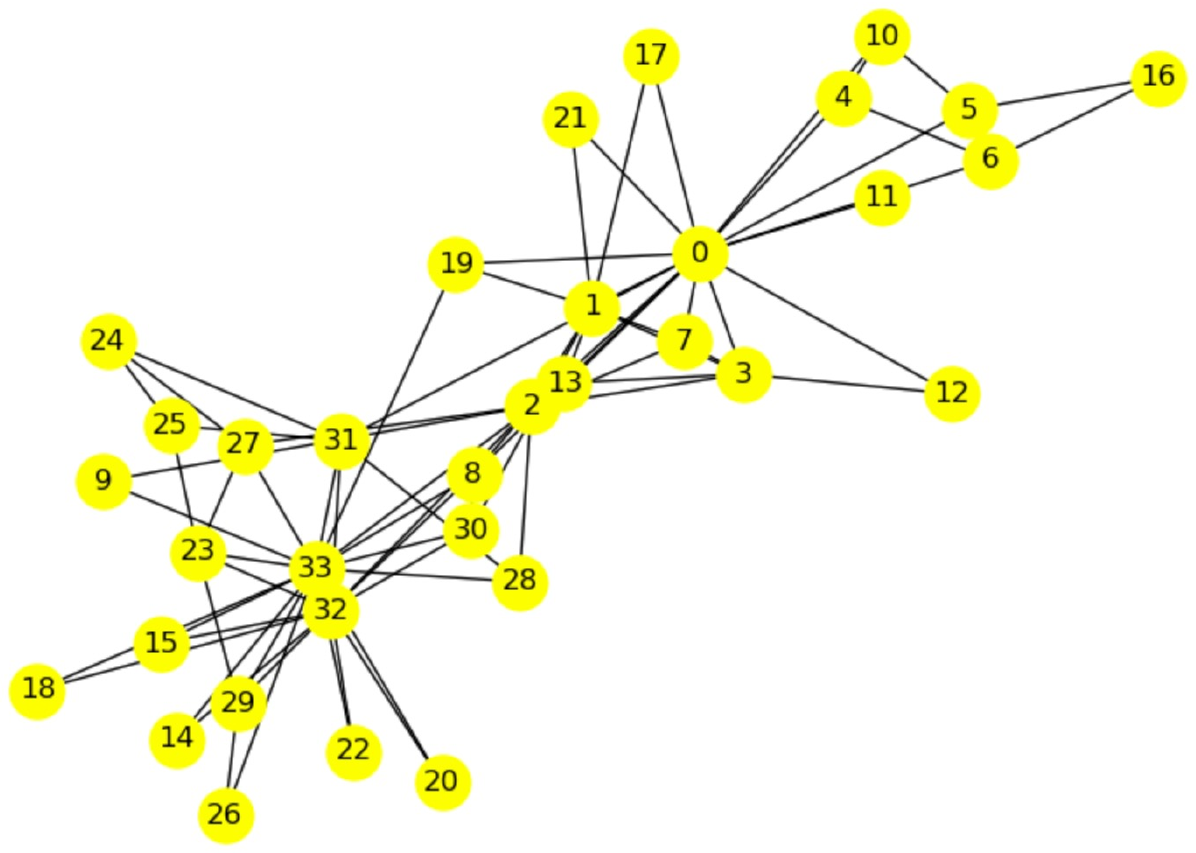}
		\caption{Illustration of Karate Club Network} 
		\label{fig:exampleproblem}
	\end{center}
 \vspace{-0.2in}
\end{figure}

For example, consider a graph, as shown in figure \ref{fig:exampleproblem}, which consists of 34 nodes and 78 undirected edges, representing frequent communication patterns among individuals in a social network. In this case, the nodes correspond to individuals, and the edges reflect the communication interactions between them. Through graph-to-text conversion, we can generate a descriptive narrative that summarizes the network’s key characteristics, such as identifying central or influential nodes, the strength of relationships between nodes (based on edge weights), and any notable community structures that emerge within the network. This process not only aids in understanding the graph's overall organization but also provides valuable insights into the social dynamics at play.

The Graph-to-text conversion will be as  follow:

\begin{mdframed}[backgroundcolor= blue!10,linewidth=1.3pt,] 
 Node 0 is connected to: 1, 2, 3, 4, 5, 6, 7, 8, \dots.\\
Node 1 is connected to: 0, 2, 3, 7, 13, 17, \dots.\\ 
\vdots  \\
Node 33 is connected to: 8, 9, 13, 14, 15, 18, \dots.
\end{mdframed}

\subsection{Community Detection via LLM Reasoning}
We propose a novel community detection method \newmethod that formulates the problem as a text-based reasoning task. The input to the language model is a natural language description of node connections in the form as given in section \ref{sec:method_GT} and the prompt to the model is:
\begin{mdframed}[backgroundcolor= blue!10,linewidth=1.3pt,] 
Graph Details + "A community is a group of nodes that are more densely connected to each other internally than to the rest of the network. You are doing community detection. Based on these node connections, which community each node belongs? Give outcome as  Node:$<$node id$>$; Community:$<$Community id$>$ format. Do not give any other text."
\end{mdframed}

The LLM implements an implicit function:

 $$f: \mathcal{T}(G) \rightarrow \mathcal{C}$$
 $$f : \text{Connectivity Descriptions} \xrightarrow{} \text{Community Assignments} $$
where 
$\mathcal{T}(G)$ denotes a textual representation of the graph (i.e., the connectivity format described above), and $C$ is the resulting community assignment. The model infers clusters by identifying patterns in the described node relationships, without explicit optimization of an objective. This approach treats the LLM as a zero-shot or few-shot reasoning engine over structural information conveyed via text, rather than a model trained explicitly on community detection tasks or graph statistics.

The output of \newmethod is captured as:
\begin{mdframed}[backgroundcolor= blue!10,linewidth=1.3pt,] 
Node:0; Community:1  \\
Node:1; Community:3  \\
\dots\\
Node:33; Community:2  
\end{mdframed}
In summary, \newmethod re-frames community detection as a language-based reasoning task, requiring no structural assumptions or predefined models. By leveraging GPT-4o's capability to interpret connectivity patterns through natural language prompts, we demonstrate a flexible and generalizable approach to discovering community structures in graph data.

%%%%%%%%%%%%%%%%%%%%%%%%%%%%%%%%%%%%%%%%%%%%%%%%%%%%%%%%%%%%%%%%
\begin{table}[t]
	\centering
	\small
	\begin{tabular}{|c||c|c|c|c|}
	\cline{1-5}
    {\bf Dataset}	& {\bf $\#V$} & {\bf $\#E$}& {\bf $\#C$} & {\bf directed} \\	\hline
	   Karate Club&$34$&$78$&$2$&N\\\hline
         Football&$115$&$613$&$12$&N\\\hline
         WebKB&$187$&$298$&$5$&Y\\\hline
        Terrorist Attacks&$645$&$3172$&$6$&Y\\\hline
         Cora&$2708$&$5278$&$7$&Y\\\hline
         CiteSeer&$3279$&$4552$&$6$&Y\\\hline
         
\hline
	\end{tabular}
	\caption{Real world social network dataset. $V$ = Number of nodes; $E$ = Number of Edges; $C$: Number of communities in the graph.}
\label{tbl:dataset} 
\end{table}
%%%%%%%%%%%%%%%%%%%%%%%%%%%%%%%%%%%%%%%%%%%%%%%%%%%%%%%%%%%%%%%%%

%% file: 040experiments.tex
\section{Experiments}
\label{obtd:experiments}
In this section, we answer following research questions:
\begin{itemize}
\item \textbf{Q1:} Does \newmethod able to provide relevant communities for different social network graphs?
\item \textbf{Q2:} Is there effect of number of nodes on \newmethod?
\item \textbf{Q3:} Does \newmethod is sensitive to different prompts?
\end{itemize}
We execute all methods for 10 times to handle hallucinations and average score is reported. 
\subsection{Dataset}
To evaluate the proposed framework at scale, we use 6 real world datasets \cite{linqsdatasets} as show in table \ref{tbl:dataset}. The dataset is publically avaiable to explore at link\footnote{\url{https://drive.google.com/drive/folders/1IKV-Qi_oz0hjAJHqCszLIbwlACmgNWXM}}.

\subsection{Baselines}
To evaluate the performance of \newmethod, we compare \newmethod with LLM based methods namely gpt-3.5-turbo \cite{openai2024gpt35}, llama3-3-70b-instruct  \cite{touvron2023llama}, claude-3.5-sonnet \cite{anthropic2024}, and gemini-1.5-pro \cite{anil2024gemini}.

\subsection{Evaluation metrics}
The evaluation metrics Normalized Mutual Information (NMI), Adjusted Rand Index (ARI), Variation of Information (VOI), and  Purity are used to evaluate the detected communities. A lower VOI value indicates a more similar clusters, while a higher VOI value suggests that the clusters are more different. For other metrics larger value implies a better algorithm performance.

\begin{figure*}
	\begin{center}
	    \includegraphics[clip,trim=0cm 3cm 0cm 3cm,width=0.45\textwidth]{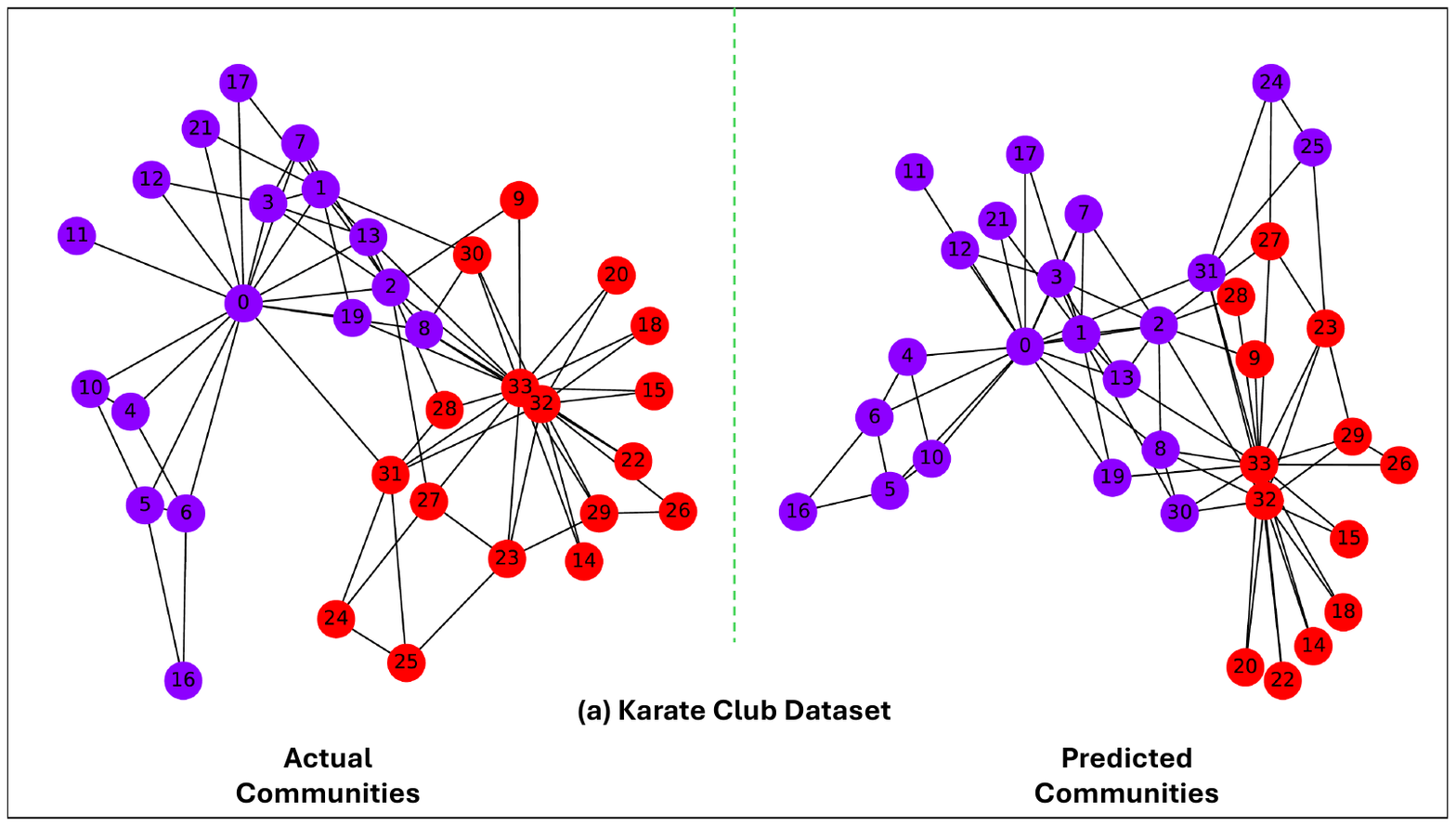}
        \includegraphics[clip,trim=0cm 3cm 0cm 3cm,width=0.45\textwidth]{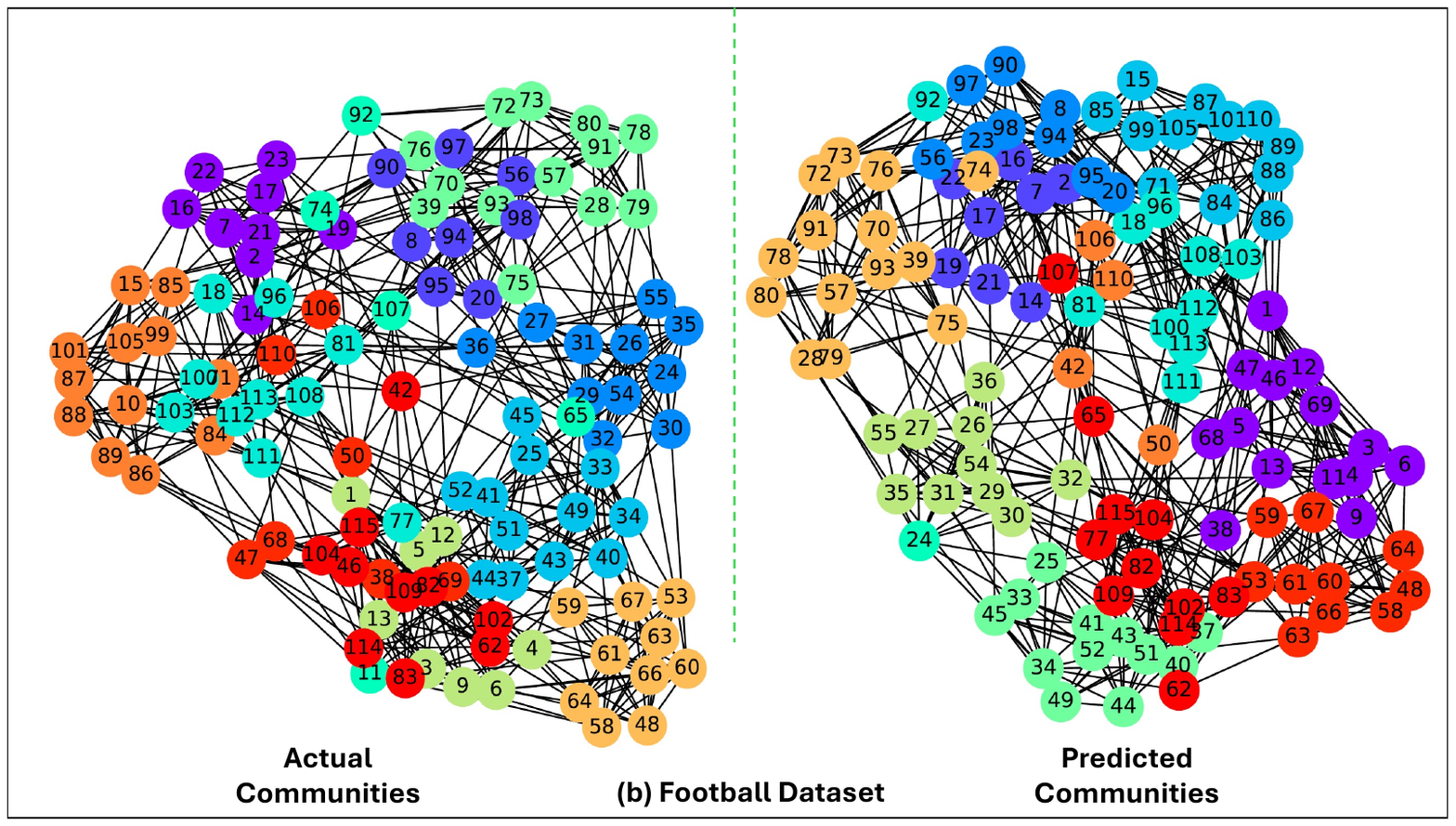}
        \includegraphics[clip,trim=0cm 3cm 0cm 3cm,width=0.45\textwidth]{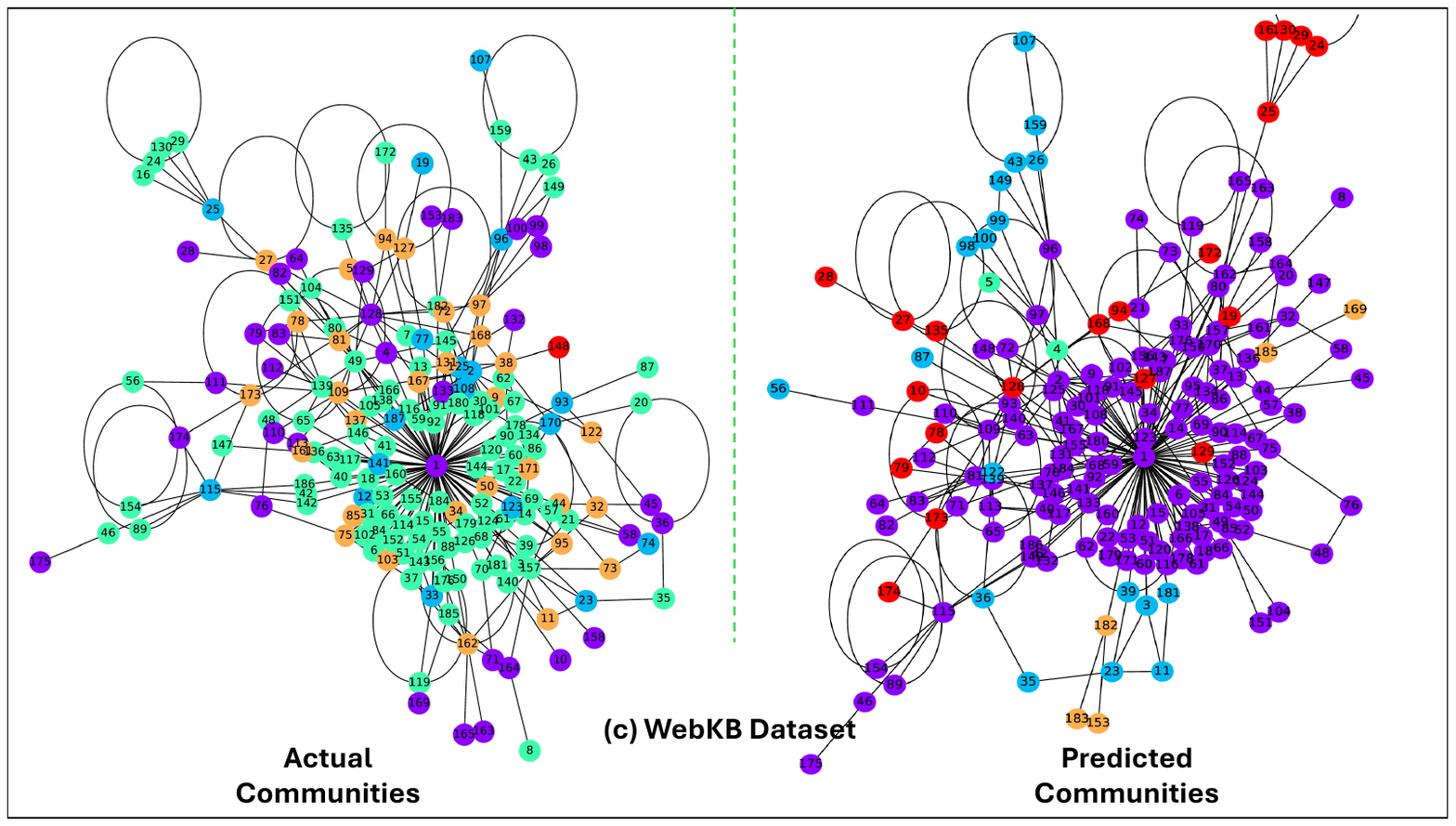}
        \includegraphics[clip,trim=0cm 3cm 0cm 3cm,width=0.45\textwidth]{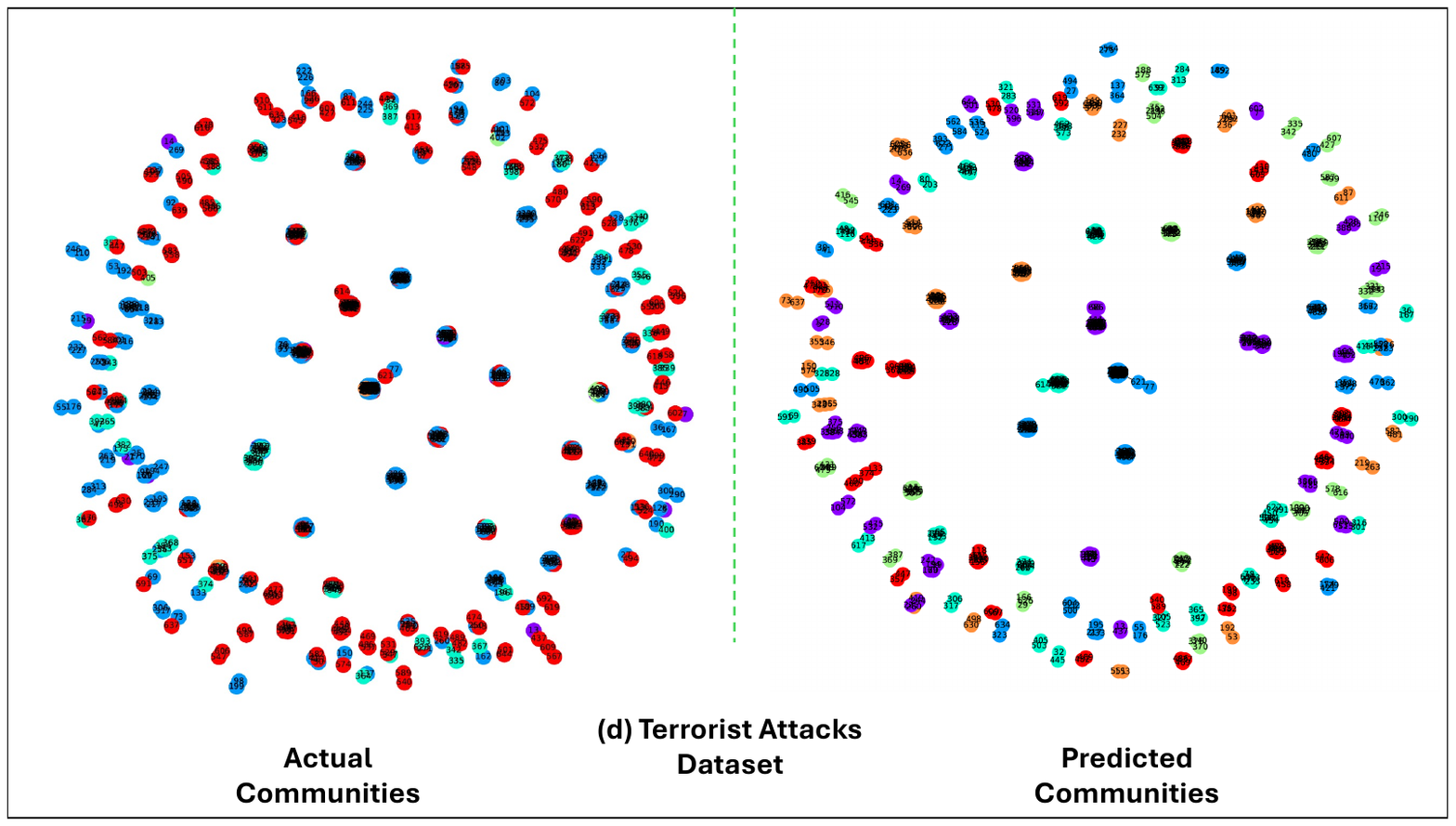}
		\caption{Illustration of community assignment by \newmethod for the graphs} 
		\label{fig:pcimage}
	\end{center}
\end{figure*}

\subsection{Experiment Results}
\subsubsection{Results on networks}
We evaluate \newmethod, our reasoning-based community detection method, across six benchmark datasets and compare its performance against leading large language models (LLMs). The results are shown in Table \ref{tbl:result_all} and Figure \ref{fig:pcimage}. For Karate Club
TribeLLM achieves NMI = $0.90$ outperforming all baselines by a significant margin (see Figure \ref{fig:pcimage}). Competing models like LLama3 and Claude struggled, with NMI values around $0.34$, indicating a poor match with ground truth. For Football dataset, \newmethod leads with NMI = $0.91$ while keeping VOI low at $0.31$. Gemini-1.5-pro was a close second in community detection quality but had a notably higher VOI ($0.40$), showing \newmethod's superior consistency. For WebKB dataset, \newmethod substantially outperforms others in both NMI ($0.38$) and ARI ($0.39$), whereas competitors score below $0.25$. This is particularly noteworthy given the relatively low ARI scores of all other models ($\leq 0.19$), highlighting proposed method's robustness in web graph structures. For Terrorist Attacks dataset, \newmethod achieves competitive scores (NMI = $0.37$, ARI = $0.34$) and maintains low VOI ($1.14$). Claude outperforms in NMI ($0.50$) and Purity ($0.73$), but \newmethod remains close while offering better balance across all metrics. For Cora and CiteSeer, \newmethod is the only model to complete inference successfully with high-quality scores (NMI = $0.35$, ARI = $0.36$). Other models either ran out of tokens (Claude, Llama3) or underperformed, indicating \newmethod's efficiency with larger graphs.

\newmethod demonstrates superior generalization and reasoning ability across diverse datasets, particularly excelling in both well-defined and noisy graph structures. It consistently outperforms baseline LLMs, many of which fail due to token limits, lack of contextual reasoning, or insufficient robustness in community assignment inference. This answer our \textbf{Q1}.
\begin{table}[t]
	\centering
	\small
	\begin{tabular}{|c||c|c|c|}
	\cline{1-4}
    {\bf Dataset}	& {\bf $\#V$} & {\bf Time Consumption}&    {\bf Input Tokens} \\	\hline
	   Karate Club&$34$&$20$& $777$\\\hline
         Football&$115$&$25$&$4,573$\\\hline
         WebKB&$187$&$41$&$3,144$\\\hline
         Terrorist Attacks&$645$&$65$&$23,618$\\\hline
         Cora&$2708$&$176$&$57,146$\\\hline
         CiteSeer&$3279$&$552$&$58,453$\\\hline

\hline
	\end{tabular}
	\caption{Effect of the number of nodes on \newmethod in terms of Time (seconds) and token used. For Cora and CiteSeer model can only predict partial nodes at a time.}
\label{tbl:tokeninfo} 
\end{table}

%%%%%%%%%%%%%%%%%%%%%%%%%%%%%%%%%%%%%%%%%%%%%%%%%%%%%%%%%%%%%%%%%
\subsubsection{Effect of the number of nodes on \newmethod}
As the number of nodes in the graph increases, the input and output token requirements for the proposed method grow proportionally as shown in table \ref{tbl:tokeninfo}, since the entire graph structure is embedded within the prompt provided to the \newmethod model. This results in larger message sizes, directly impacting memory consumption and model processing capacity. Additionally, increased token length leads to longer inference times due to the model's attention mechanism scaling with input size. These factors collectively introduce computational overhead, making scalability a key consideration. Efficient graph encoding or summarization strategies to mitigate these effects in larger graphs will be part of our future research work and is out of scope for this work. This answer our \textbf{Q2}.
\begin{table}[h]
	\centering
	\small
    \begin{tabular}{|c||c|c|c|c|c|}
   \cline{1-6}
    \multicolumn{6}{|c|}{\bf Karate Club Dataset}\\
    \hline
    {\bf Method}	& {\bf NMI} &  {\bf ARI}&     {\bf Purity} &     {\bf VOI}&     {\bf Input Tokens}\\	\hline
	 Prompt 1 &$0.78$&$0.84$& $0.90$&$0.22$&$778$\\\hline
      Prompt 2&$0.64$&$0.68$& $0.75$&$0.56$&$770$\\\hline
     Prompt 3&$0.55$&$0.63$& $0.66$&$0.85$&$793$\\\hline
       \textbf{\newmethod}&$0.90$&$0.93$& $0.98$&$0.11$&$777$\\\hline
 \hline
	\end{tabular}
\caption{Performance of \newmethod w.r.t different variation of prompts.}
\label{tbl:result_prompt} 
\end{table}
\subsubsection{Prompt comparisons}
We performed experiment on Football dataset using 3 different version of \newmethod prompt as given below. 
\begin{itemize}
    \item  {\bf Prompt 1}: A community is a group of nodes that are more densely connected to each other internally than to the rest of the network. Based on these node connections, output in the format Node:$<node id>$; Community:$<Community id>$.
 \item  {\bf Prompt 2}: You are doing community detection. Assign each node to a community. Give outcome as Node:$<node id>$; Community:$<Community id>$ format. Do not give any other text.
 \item {\bf Prompt 3}: A community is a group of nodes that are more densely connected to each other internally than to the rest of the network. You are doing community detection. Along with assigning communities, provide a brief justification for each decision. Format as Node:$<node id>$; Community:$<Community id>$; Reason:$<reason>$.
\item  {\bf Prompt 4} (\newmethod): A community is a group of nodes that are more densely connected to each other internally than to the rest of the network. You are doing community detection. Based on these node connections, which community each node belongs? Give outcome as  Node:$<$node id$>$; Community:$<$Community id$>$ format. Do not give any other text.
\end{itemize}

Table \ref{tbl:result_prompt} shows the results of this experiment. The first prompt doesn’t explicitly say the task is community detection, leaving the intent unclear and missing critical information. However, LLM was able to stick to the community detection work. The second prompt is incomplete and lacks the actual node connections or network structure to base decisions on. The proposed method performance degrades with incomplete information. The third prompt is closer to actual ask but does not ask for “no extra text” constraint and introduces justification, which diverted the model from a concise mapping. The performance is lower of all the prompt for this. The forth prompt gives the best results and we choose this as our model's prompt for evaluating the other datasets. As we know that language models are highly sensitive to wording, context, and phrasing. variation in inference is expected behavior, as these models rely on learned statistical patterns and subtle variations can shift the model's interpretation or emphasis. This answer our \textbf{Q3}.

%%%%%%%%%%%%%%%%%%%%%%%%%%%%%%%%%%%%%%%%%%%%%%%%%%%%%%%%%%%%%%%%
\begin{table*}[t]
	\centering
	\small
    \begin{tabular}{|c||c|c|c|c|}
   \cline{1-5}
    \multicolumn{5}{|c|}{\bf Karate Club}\\
    \hline
    {\bf Method}	& {\bf NMI} &  {\bf ARI}&     {\bf Purity} &     {\bf VOI}\\	\hline
	gemini-1.5-pro &$0.84$&$0.88$& $0.97$&$0.16$\\\hline
        gpt-3.5-turbo&$0.82$&$0.85$& $0.87$&$0.23$\\\hline
        llama3-3-70b-instruct&$0.34$&$0.26$& $0.76$&$0.62$\\\hline
        claude-3.5-sonnet&$0.34$&$0.27$& $0.77$&$0.61$\\\hline
        \textbf{\newmethod}&$0.90$&$0.93$& $0.98$&$0.11$\\\hline
 \hline
	\end{tabular}

\begin{tabular}{|c||c|c|c|c|}
   \cline{1-5}
    \multicolumn{5}{|c|}{\bf Football}\\
    \hline
    {\bf Method}	& {\bf NMI} &  {\bf ARI}&     {\bf Purity} &     {\bf VOI}\\	\hline
	gemini-1.5-pro &$0.87$&$0.82$& $0.87$&$0.40$\\\hline
        gpt-3.5-turbo&$0.52$&$0.28$& $0.48$&$1.37$\\\hline
        llama3-3-70b-instruct&$0.54$&$0.32$& $0.57$&$1.54$\\\hline
        claude-3.5-sonnet&$0.53$&$0.31$& $0.53$&$1.55$\\\hline
        \textbf{\newmethod}&$0.91$&$0.84$& $0.91$&$0.31$\\\hline
 \hline
	\end{tabular}

        \begin{tabular}{|c||c|c|c|c|}
   \cline{1-5}
    \multicolumn{5}{|c|}{\bf WebKB}\\
    \hline
    {\bf Method}	& {\bf NMI} &  {\bf ARI}&     {\bf Purity} &     {\bf VOI}\\	\hline
	gemini-1.5-pro &$0.24$&$0.18$& $0.57$&$1.42$\\\hline
        gpt-3.5-turbo&$0.25$&$0.19$& $0.58$&$1.29$\\\hline
        llama3-3-70b-instruct&$0.16$&$0.12$& $0.55$&$1.22$\\\hline
        claude-3.5-sonnet&$0.10$&$0.11$& $0.58$&$1.15$\\\hline
        \textbf{\newmethod}&$0.38$&$0.39$& $0.67$&$0.89$\\\hline
 \hline
	\end{tabular}

        \begin{tabular}{|c||c|c|c|c|}
   \cline{1-5}
    \multicolumn{5}{|c|}{\bf Terrorist Attacks}\\
    \hline
    {\bf Method}	& {\bf NMI} &  {\bf ARI}&     {\bf Purity} &     {\bf VOI}\\	\hline
	gemini-1.5-pro &$0.33$&$0.32$& $0.53$&$2.04$\\\hline
        gpt-4-turbo&$0.35$&$0.31$& $0.50$&$1.14$\\\hline
        llama3-3-70b-instruct&$0.41$&$0.30$& $0.61$&$2.34$\\\hline
        \textbf{claude-3.5-sonnet} &$0.50$&$0.36$& $0.73$&$1.12$\\\hline
        \newmethod&$0.37$&$0.34$& $0.56$&$1.14$\\\hline
 \hline
	\end{tabular}

        \begin{tabular}{|c||c|c|c|c|}
   \cline{1-5}
    \multicolumn{5}{|c|}{\bf Cora}\\
    \hline
    {\bf Method}	& {\bf NMI} &  {\bf ARI}&     {\bf Purity} &     {\bf VOI}\\	\hline
	gemini-1.5-pro &$0.23$&$0.13$& $0.41$&$1.58$\\\hline
        gpt-4-turbo&$0.25$&$0.27$& $0.54$&$1.23$\\\hline
       llama3-3-70b-instruct&$-$&$-$& $-$&$-$\\\hline
        claude-3.5-sonnet&$-$&$-$& $-$&$-$\\\hline
        \textbf{\newmethod}&$0.35$&$0.36$& $0.56$&$1.18$\\\hline
 \hline
	\end{tabular}

        \begin{tabular}{|c||c|c|c|c|}
   \cline{1-5}
    \multicolumn{5}{|c|}{\bf CiteSeer}\\
    \hline
    {\bf Method}	& {\bf NMI} &  {\bf ARI}&     {\bf Purity} &     {\bf VOI}\\	\hline
        gemini-1.5-pro &$0.18$&$0.12$& $0.23$&$1.88$\\\hline
        gpt-4-turbo&$0.15$&$0.17$& $0.38$&$1.53$\\\hline
       llama3-3-70b-instruct&$-$&$-$& $-$&$-$\\\hline
        claude-3.5-sonnet&$-$&$-$& $-$&$-$\\\hline
        \textbf{\newmethod}&$0.21$&$0.22$& $0.38$&$1.32$\\\hline
 \hline
	\end{tabular}
	\caption{Performance of \newmethod w.r.t state-of-art methods. Bold letter shows which method outperformed for the dataset. Symbol $-$ indicates model was out of tokens for full data and performance in terms of NMI was low ($<0.1$). The performance variance for each dataset is between $\pm 0.01$ and $\pm 0.03$.}
\label{tbl:result_all} 
\end{table*}
%%%%%%%%%%%%%%%%%%%%%%%%%%%%%%%%%%%%%%%%%%%%%%%%%%%%%%%%%%%%%%%%%

%% file: 050conclusions.tex
\section{CONCLUSIONS AND FUTURE WORK}
\label{sec:conclusions}
In conclusion, our proposed method \newmethod leverages the capabilities of GPT-4o to perform effective community detection in social network graphs through a two-step process: graph-to-text conversion followed by LLM-based reasoning. Traditional LLMs struggle with directly processing large and complex graphs due to input size limitations and lack of native graph understanding. Our approach addresses this by designing a structured conversion technique and tailored prompts that enable the LLM to capture both the semantics and topology of the graph. Experimental results confirm that this method significantly enhances the model's ability to understand graph structures and accurately detect communities. For future work, we aim to extend this framework to handle much larger and dynamic graphs, pushing the boundaries of scalable community detection using language models. Overall, this study contributes a novel methodology for incorporating LLM reasoning into structured graph tasks, offering practical guidance and foundational understanding for future work at the intersection of natural language processing and graph analytics.